# Unconventional charge density wave in Kagome metal BaFe$_2$Al$_9$

Liucheng Chen[1], Mingwei Ma[1*], Xiaohui Yu[1,2*], Fang Hong[1,2*]

[1] *Beijing National Laboratory for Condensed Matter Physics, Institute of Physics, Chinese Academy of Sciences, Beijing 100190, China*

[2] *School of Physical Sciences, University of Chinese Academy of Sciences, Beijing 100190, China*

Corresponding authors: mw_ma@iphy.ac.cn; yuxh@iphy.ac.cn; hongfang@iphy.ac.cn

**The charge density wave (CDW) is a macroscopic quantum state characterized by long-range lattice distortion and modulated charge density. Conventionally, CDWs compete with other electronic orders (e.g., superconductivity) and are suppressed under hydrostatic pressure. Intriguingly, the Kagome-variant metal BaFe$_2$Al$_9$, crystallized in a three-dimensional structure, exhibits pressure-enhanced CDW ordering, where the transition temperature (T$_{CDW}$) rises from ~110 K to room temperature near 3.6 GPa. The lattice structure was checked by both powder and single crystal x-ray diffraction (XRD). The XRD data reveals an abnormal lattice expansion along *a* axis near 4-5 GPa upon compression. The strongly suppressed diffraction intensity and splitting diffraction spots from single crystal indicates cracking and breakdown to smaller pieces, indicative of an intrinsic first-order transition character. This anomalous response implies a CDW mechanism dominated by electron-electron and/or electron-phonon correlations, distinct from Fermi-surface nesting in conventional systems. Concomitant dome-shaped pressure-dependent resistance suggests competing electronic phases. Our work establishes BaFe$_2$Al$_9$ as a 3D Kagome platform with unconventional CDW behavior and strong electron-phonon coupling, which provides an alternative platform to explore the electron correlation induced exotic electronic states and other potential emergent quantum phenomenon.**

**Introduction**

Charge density wave (CDW) is a unique charge ordering due to the structural instability and consequent long-range lattice distortion, which helps to lower the energy of the compounds [1]. Generally, such an ordering exists in low-dimensional materials, especially in quasi-one-dimensional compounds, such as $NbS_3$ and $HfTe_3$ [2,3]. Surprisingly, $NbS_3$ even undergoes three steps of CDW transition [2]. The two-dimensional transition metal dichalcogenides are also typical platforms showing the CDW, such as $1T-TaS_2$ and $2H-NbSe_2$ [4-6]. However, the charge density wave is rarely observed in three dimensional systems. The CDW may force the compounds transform from metal to an insulator [4], though it only opens gap at certain momentum near the Fermi surface in most cases. The correlation and competition between CDW and other quantum ordering, such as superconductivity, has been an important research topic in condensed matter physics community for decades [7]. The CDW has also been observed in unconventional superconductors, such as the cuprates and iron-based superconductors [8,9], and even in intensely studied nickelates in latest a few years [10,11]. Kagome lattice compounds have recently attracted significant research interest due to their unique ability to simultaneously host multiple quantum phenomena - including CDW, van Hove singularities, Dirac cones, and flat electronic bands [12]. The most studied compound is the $CsV_3Sb_5$ and its sister compounds [13,14]. Multiple CDW transitions have been observed in $CsV_3Sb_5$ by nuclear magnetic resonance (NMR) study under pressure, together with a complex double dome-shape superconductivity phase diagram [15,16]. More recently, FeGe has been reported to be a rare three dimensional Kagome system with both antiferromagnetism and CDW [17], different from the non-magnetic $CsV_3Sb_5$. The magnetic exchange splitting in FeGe drives the van Hove singularities near the Fermi level and consequent CDW formation, and it demonstrates a good system showing complex electron correlation and intertwined magnetism and CDW [18]. Therefore, the Kagome lattices have been an ideal platform to explore the electron correlation among CDW, magnetism, van Hove singularities, Dirac electrons, and flat bands, which can produce various exotic quantum states [13,14].

$BaFe_2Al_9$, a Kagome variant with typical three dimensional structure rather than a weak linked layered structure, was reported to host CDW between 80 and 110 K, based on resistivity measurement and transmission electron microscopy characterization [19]. Meanwhile, previous theoretical calculation demonstrates the existence of van Hove singularities, Dirac cone, and flat bands near Fermi surface [19]. Then, this compound comes to our attention for potential superconductivity, providing that the CDW could be suppressed by pressure. With this motivation, we synthesized $BaFe_2Al_9$ single crystal, did the electrical transport measurement and examined the crystal structure evolution by both powder and single crystal x-ray diffraction. However, the results are far away from what we expected, the CDW was not suppressed but greatly enhanced from the initial ~110 K to room temperature. In this case, the superconductivity was not observed. Such a behavior is quite different from the conventional CDW systems with pressure-suppression behavior. Here, $BaFe_2Al_9$ behaviors as an unconventional CDW system, which is quite rare. Meanwhile, the transport measurement demonstrates that the resistance was enhanced as well due to the enhanced CDW behavior. The XRD-based structure analysis reveals a first-order phase transition with abnormal pressure-driven lattice parameter expansion along *a* axis, suggesting a strong electron-phonon coupling effect during the CDW transition. This work not only reports a new CDW system with unconventional feature in form of a three-dimensional structure, but also demonstrates that

BaFe$_2$Al$_9$ would be a unique platform to explore the electron correlation and potential superconductivity by strain engineering, such as epitaxial thin film or chemically doped bulk crystal with tensile strain. Meanwhile, a few open questions are still waiting for answers in near future: 1) what is the exact lattice structure in CDW phase? 2) What is the mechanism of pressure-enhanced CDW? 3) What is the mechanism of pressure dependent dome-shaped resistance at room temperature?

**Experiment**

1) **Single crystal synthesis**

   The BaFe$_2$Al$_9$ single crystals were grown from an aluminum-rich melt, following the similar procedure in previous work [19]. High purity barium pieces, iron powder, and aluminum pieces from Alfa Aesar were loaded in an alumina crucible at the atomic composition of Ba:Fe:Al = 6:8:86, and then were sealed in a fused silica ampoule with argon filling protection. The mixture was heated to 1150 °C lasting 24 hours to make sure homogeneous reaction, and then slowly cooled down to 1000 °C for 10 days, during which the crystals started to grow. The Al solution was removed by taking the silica ampoule from the hot furnace and consequent centrifuge treatment.

2) **High-pressure structural characterizations.**

   a) **Synchrotron powder x-ray diffraction under pressure**

   The structure is examined by the synchrotron x-ray diffraction at Beamline 15U in Shanghai Synchrotron Research facility (SSRF, China). The x-ray wavelength is 0.6199 Å and beam size is about 2*4 μm$^2$. The powder sample was obtained by grinding the single crystal, making sure a pure phase at ambient condition. The powder sample was loaded into the sample chamber with diameter of ~ 150 μm and compressed between diamond culets of 300 μm diameter. CeO$_2$ was used to calibrate the instrument parameters including sample to detector distance, beam center and detector tilt. Silicon oil was used as the transmitting medium. The ruby fluorescence method [20] was used to determine the pressure.

   b) **Single crystal x-ray diffraction under pressure at lab-based diffractometer**

   High pressure single crystal x-ray diffraction (XRD) was carried out on a D8 VENTURE x-ray diffractometer with a high intensity Mo x-ray source and a PHOTON II CPAD area detector. The x-ray was collimated to a beam size of ~75 μm in diameter and focused on the sample. A small piece of single crystal was loaded into the sample chamber with diameter of ~ 150 μm in a diamond anvil cell with culets of 300 μm diameter. The system has been calibrated by using CeO$_2$ as the standard material. Neon gas was used as the transmitting medium to achieve a good hydrostatic environment. The ruby fluorescence method [20] was used to determine the pressure.

3) **High-pressure transport property measurements**

   The standard four-probe electrical resistance measurement under high pressure was carried out in a commercial cryostat from 1.7 K to 300 K by a Keithley 6221 current source and a 2182A nanovoltmeter. A BeCu alloy diamond anvil cell (DAC) with two opposing anvils was used to generate high pressure. In current experiments, a thin single crystal sample cleaved from the bulk crystal was loaded into the sample chamber in a rhenium gasket with c-BN insulating layer and silicone oil as pressure medium. A ruby ball is loaded to serve as internal pressure standard [20].

**Results and discussion**

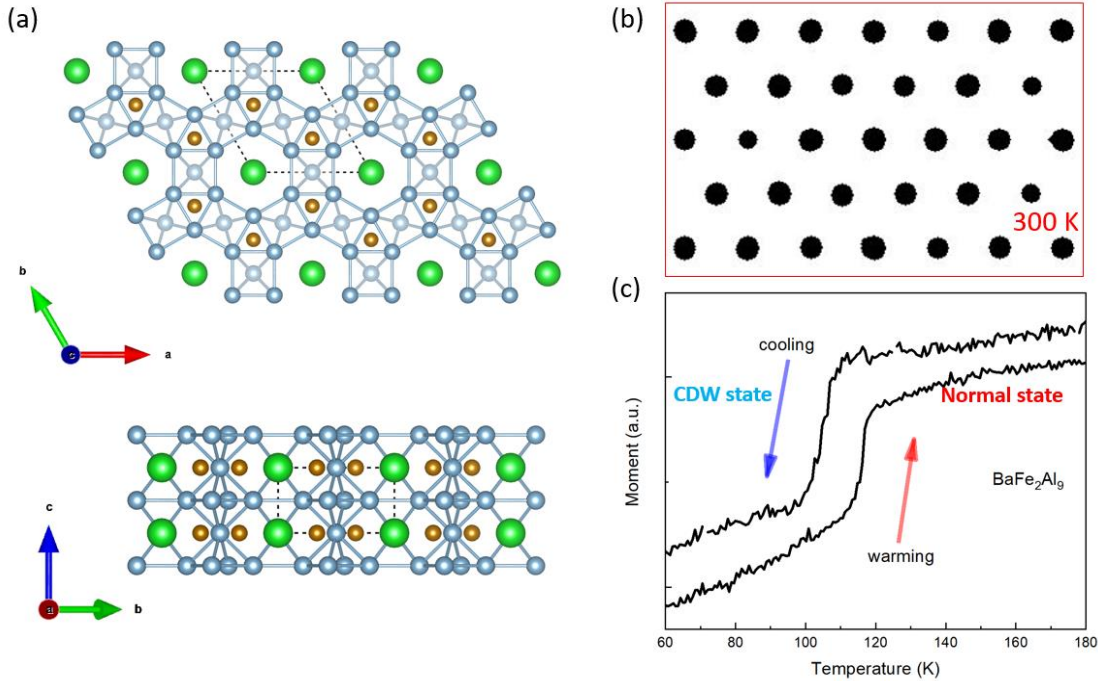

Fig.1 The atomic structure of Kagome metal $BaFe_2Al_9$ at room temperature and the first-order charge density wave phase transition near 110 K. (a) The atomic structure along *c* and *a* axis, respectively. The green atoms: Ba; brown atoms: Fe; shallow blue atoms: Al. (b) The typical single crystal x-ray diffraction, projection along *c**, showing a hexagonal phase (Space group: No. 191, P6/mmm). (c) The magnetic response near the CDW transition with an obvious hysteresis behavior.

The $BaFe_2Al_9$ single crystal was obtained by the melting method with Al solution. The crystal sizes were about 1-3 mm in diameter without well-defined shape. It is easy to get a shiny surface by cutting the bulk crystal along the cleave direction. To check the crystal quality, we did some basic characterization before high pressure experiment. Firstly, the crystal was examined by the single crystal diffractometer and a set of nice data was obtained, as seen in Fig. 1(b), and the diffraction spots are projected along *c**. The crystal was confirmed to be in form a P6/mmm structure, and detailed structure information is listed in Table 1. Then, to check the CDW transition quickly, we measured the magnetic response of $BaFe_2Al_9$ single crystal, as it was expected to show a drop of magnetic moment near the CDW transition upon cooling [19]. The testing field was set to 2000 Oe. The overall magnetic response is weak because of the non-magnetic behavior in $BaFe_2Al_9$. Clearly, we did see the magnetic moment drop near 110 K, and a hysteresis behavior was observed between the cooling and warming processes. This behavior is almost consisted with previous work [19], apart from a ~10 K difference on CDW transition temperature.

**Table 1 the structure information for BaFe$_2$Al$_9$ single crystal at ambient condition**

| Lattice parameters (Å) | a | c | V (Å$^3$) |
|---|---|---|---|
| P6/mmm (No.19) | 8.0134 | 3.9337 | 218.759 |
| **Atom** | x | y | z |
| Ba | 0 | 0 | 0 |
| Fe | 0.6667 | 0.3333 | 0 |
| Al-1 | 0.7853 | 0.5705 | 0.5 |
| Al-2 | 0.5 | 0.5 | 0 |

To study the CDW and potential superconductivity, electrical transport measurement was carried out under pressure. A tiny piece of sample was cleaved from the bulk crystal and standard four-probe geometry was used to place the electrodes, making sure a good data collecting stability. The pristine transport results are displayed in Fig. 2(a-b). At low pressure, a kink was seen near 125 K at 0.6 GPa as presented in Fig. 2(a), indicative of a CDW transition. Unexpectedly, it became much more pronounced with elevated transition temperature under higher pressure, contrary to the suppression behavior in most CDW systems. In current work, the kink at low pressure is less sharp than that in previous work [19], and it should be due to the sample direction difference. This also indicates a strong anisotropy behavior of the CDW ordering in this three-dimensional compound. The CDW transition temperature (T$_{CDW}$) increases quickly upon further compression and exceeded room temperature at 3.8 GPa. By plotting the relationship between the T$_{CDW}$ and pressure, it is estimated that the T$_{CDW}$ is 300 K at ~3.6 GPa, as presented in Fig. 2(c). Meanwhile, it is also noticed that the sample's resistances at room temperature or low temperature always increases with pressure up to 10.4 GPa as seen in Fig. 2(b), suggesting an enhanced electron localization, which is consistent with the enhanced T$_{CDW}$. Such an abnormal electron localization and T$_{CDW}$ may be related to the strong electron correlation behavior prevailing in Kagome lattice. Previous calculation on the band structure of BaFe$_2$Al$_9$ at ambient pressure demonstrates the presence of van Hove singularity, Dirac cone, and even flat band near the Fermi surface, and these unique electron states allow strong electron correlation/interaction [19]. In this case, it is reasonable to assign the pressure-driven abnormal behavior to the strong electron correlation effect. Due to the largely enhanced T$_{CDW}$, we failed to observe any signature of superconductivity. Above 10.4 GPa, the resistance at room temperature started to decrease with pressure, while the resistance at 2 K was still higher than that at 10.4 GPa. Further compression to 14.1 GPa, the resistances over the whole temperature range decreased. The dome-shape resistance under pressure, as presented in Fig.2(d), indicates a competition behavior of electron correlated states as mentioned above, which is worthy further study.

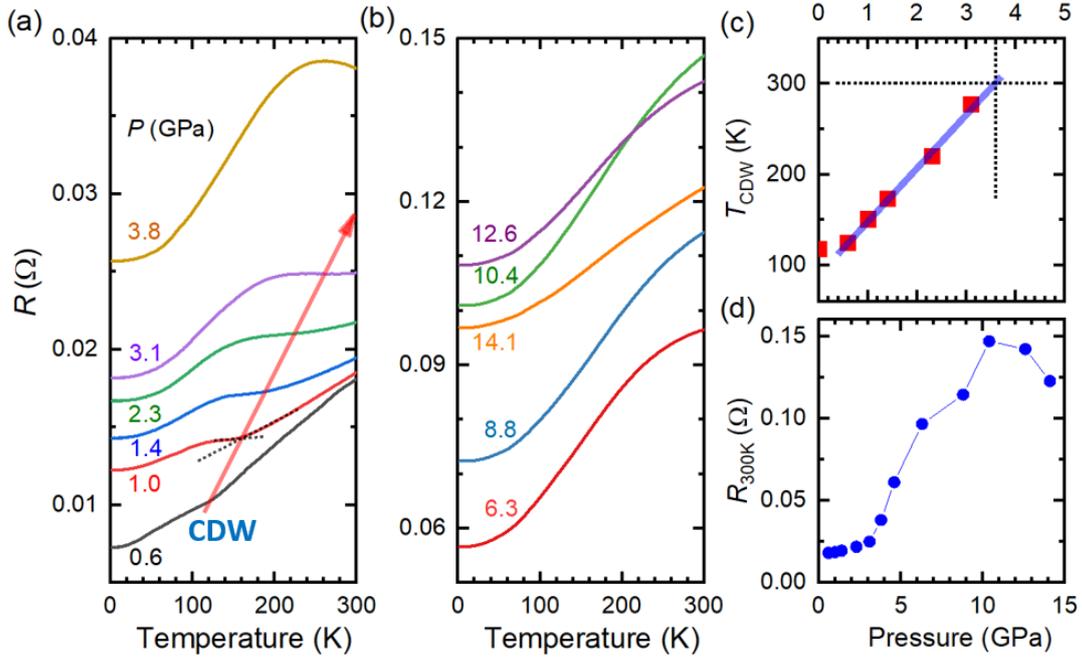

Fig.2 The transport measurement on the charge density wave transition in BaFe$_2$Al$_9$ under pressure. (a) The R-T curves from 0.6 to 3.8 GPa. (b) The R-T curves from 6.3 to 14.1 GPa. (c) The extracted CDW temperatures, suggesting a CDW temperature of 300 K near 3.6 GPa. (d) The pressure dependent resistance of BaFe$_2$Al$_9$ sample at 300 K.

To track the structural change under pressure and try to understand the unconventional CDW behavior, we examined the structure evolution of BaFe$_2$Al$_9$ by both powder x-ray diffraction and single crystal diffraction. Fig. 3 presents the powder XRD patterns and lattice parameters from structural refinement. As shown in Fig. 3(a), BaFe$_2$Al$_9$ almost keeps the original P6/mmm phase and it is hard to see clear signature of any structural phase transition. Since the T$_{CDW}$ was enhanced to 300 K at ~3.6 GPa, as mentioned above, it is reasonable to expect some super-structure feature with extra diffraction from the CDW state. However, the experimental data did not show such a kind of diffraction peaks. According to previous study on BaFe$_2$Al$_9$ by low temperature transmission electron microscopy [19], the BaFe$_2$Al$_9$ crystal sample will break into smaller pieces due to the large lattice strain during the CDW transition. It results in a weak and catastrophic electron diffraction patters consequently, the data cannot solve the exact crystal structure of the CDW state, and extra neutron scattering experiment also failed to solve this problem [19]. Hence, it is not strange that the powder x-ray diffraction failed to reveal the super-lattice generated by the CDW, due to both the weak CDW diffraction itself and crystal breaking. In spite of this unsolvable problem, there is some positive news which can help to identify whether the CDW-phase structure is presented at room temperature above 3.6 GPa. The big difference between the normal state and the CDW state is their lattice parameters. By refining the x-ray patterns, we extracted the lattice parameter by using the same space group. As displayed in Fig.3 (b), we do observe some abnormal behavior in pressure dependent $a$ values. Between 4 and 5 GPa, the $a$ values did not follow the overall decreasing trend under pressure, but showed an expansion. Such a behavior is consisted with that observed in low-temperature experiment at ambient pressure, in which $a$ value increased when temperature was cooling down below the T$_{CDW}$ [19]. However, the expansion ratio is only 0.161%, much smaller than the ratio of 0.5% under ambient pressure [19]. Meanwhile, the $c$ value decreased by about 0.244% as seen in Fig. 3(c), which is also smaller than the ratio of 1.5% under ambient pressure [19]. Such a structural change can be displayed in a much clear way by plotting the pressure dependent a/c value, as seen in Fig. 3(d). It is obvious that the a/c ratio shows a different pressure dependence below and above 4-5 GPa. The critical pressure for CDW transition ($P_{CDW}$) here is 4-5 GPa, while it was supposed to be 3.6 GPa based on transport results. It is accepted that the electronic response is generally stronger at the early stage of a structural phase transition while the XRD characterization is less sensitive, especially for such a first-order transition. Since the change of $a$ and $c$ value is small near $P_{CDW}$, the volume change is not so strong and even unnoticeable, and therefore the P-V curve can be well fitted by a same set of equation of state, as presented in Fig. 3(e).

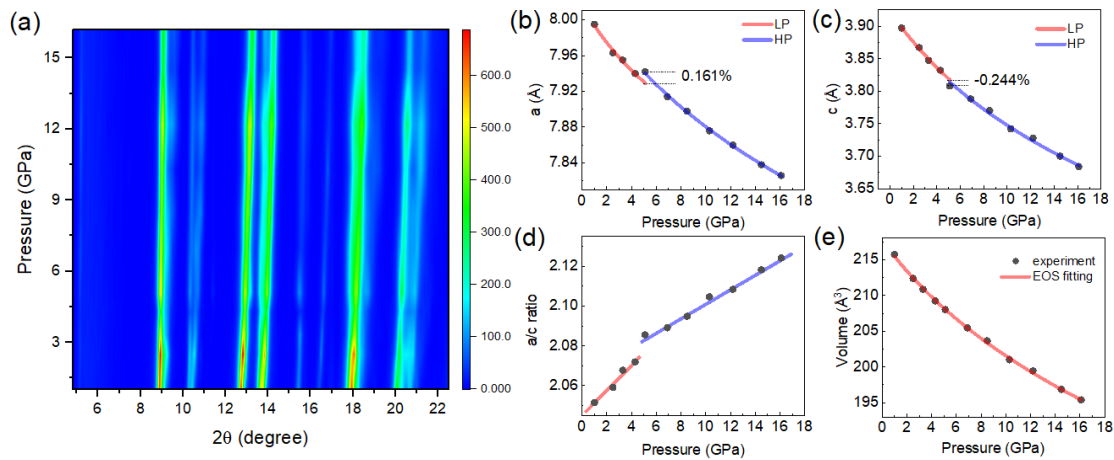

Fig.3 The powder x-ray diffraction of BaFe$_2$Al$_9$ under pressure. (a) The 2D contour plot for x-ray diffraction patterns up to ~16 GPa. (b-e) The lattice parameters obtained from the refinement.

Considering the small change in lattice parameters from synchrotron x-ray diffraction, which may be caused by experimental uncertainty/error or mistake, we did single crystal x-ray diffraction up to 6.7 GPa, making sure the data reliability. As given in Fig. 4(a), the diffraction at 0.3 GPa from BaFe$_2$Al$_9$ single crystal shows many perfect diffraction spots, confirming a good crystalline quality. With increasing pressure, the diffraction spot becomes distorted or splitting, especially above 5 GPa. Fig. 4(b) presents the diffraction pattern projected along $a^*$ at 6.7 GPa: clear distortions can be found in various spots and some spots lose the intensity strongly, compared with those at 0.3 GPa. The distortion and splitting are typical signal for strong strain or/and crystal cracking. It has to be noted that neon gas was used as pressure medium in the single crystal XRD experiment, which provides a very good hydrostatic pressure environment at least to ~15 GPa[21]. Hence, the crystal cracking/breaking is a result of CDW phase transition itself rather than non-hydrostaticity. The single crystal XRD data were analyzed by the Olex2 software embedded in the diffractometer facility. The lattice parameters are obtained and presented in Fig. 4(c-d). Again, we see the lattice expansion along $a$ axis near 5 GPa. This experiment further confirms the presence of CDW transition between 4 and 5 GPa, together with the strong lattice strain and consequent crystal breaking. The other motivation to carry out single crystal diffraction is to search for possible superlattice diffraction from the CDW, since single crystal generally has a much higher diffraction signal than powder diffraction. However, from the data presented in Fig. 4(a) and 4(b), we can't see any feature of CDW superlattice. One reason for the failure is the low diffraction intensity from CDW, and the other is the strain-induced crystal breaking. In this case, to figure out the exact lattice structure in CDW state, further work is still required. The synchrotron-based nano-beam single crystal diffraction may help, combined with laser heating or AI-assisted structure simulation and searching.

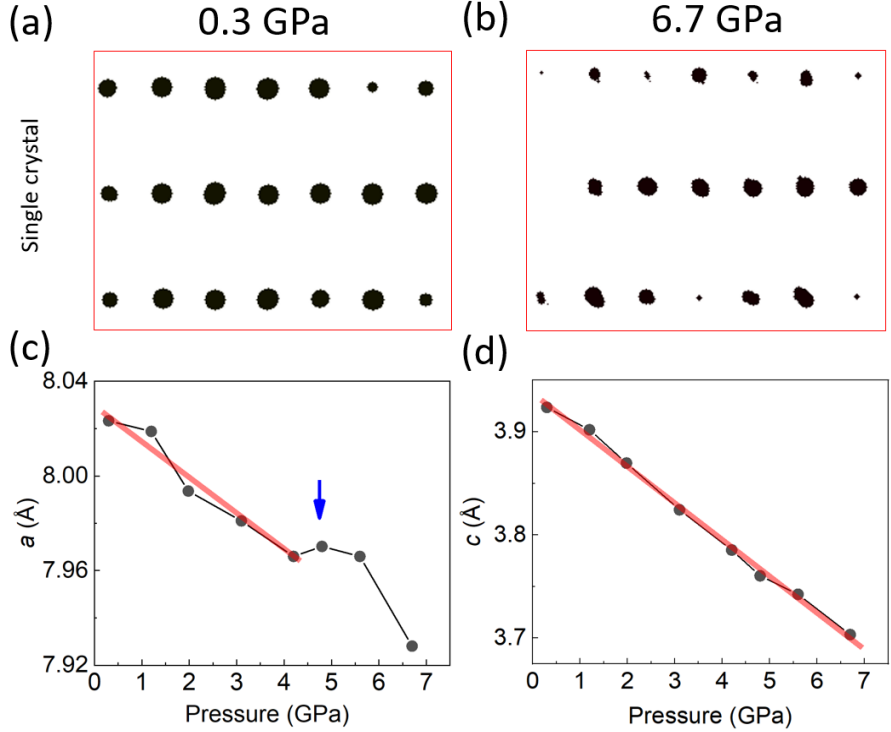

Fig. 4 The single crystal x-ray diffraction and lattice parameters of $BaFe_2Al_9$ under pressure. (a-b) The single crystal diffraction projection along $a*$ at 0.3 and 6.7 GPa, respectively. (c-d) The pressure dependent lattice parameters. At 6.7 GPa, the diffraction spots show clear distortion and even splitting with reduced intensity, indicative of crystal breaking (due to the abnormal expansion of lattice parameter $a$).

The unconventional CDW is quite rare, and was only observed in a few systems. Two typical examples are the unconventional CDW found in non-Kagome $SmNiC_2$ and Kagome FeGe [22,23]. In $SmNiC_2$, the enhanced CDW behavior is commensurate with the suppressed ferromagnetism, suggesting a competing behavior between the CDW and magnetism [22]. The resistance increases as well in $SmNiC_2$ when the CDW is enhanced [22]. In case of FeGe, the CDW is enhanced under pressure but the resistance decreases, while another FeGe sample shows pressure-enhanced resistance but there is no CDW observed [23]. The unconventional CDW in FeGe was claimed to be driven by the delicate balance between magnetic energy savings and structural energy costs by Ge1 dimerization, excluding the mechanism of van Hove singularities' nesting [23], which is also supported by another theoretical work (they claim the antiferromagnetic coupling plays a critical role in CDW formation) to some extent [24]. Earlier study on FeGe claims that CDW arises from the combination of electron-correlations-driven AFM order and van Hove singularities-driven instability. In spite of some debates existing, it still comes a consensus that the CDW in FeGe is correlated with the antiferromagnetism. It is noted that the magnetic response during the CDW transition in FeGe also shows large hysteresis and the initial CDW is also about 110 K, though CDW in FeGe is much less sensitive to external pressure [23]. In $BaFe_2Al_9$, it is expected to be paramagnetic [19], rather than ferromagnetic in $SmNiC_2$ or antiferromagnetic in FeGe. In this case, a delicate study on the mechanism of the CDW in $BaFe_2Al_9$ is still required, and it may stem from a new mechanism.

## Conclusion

The charge density wave in Kagome metal BaFe$_2$Al$_9$ is studied by pressure engineering. Unexpectedly, the CDW temperature is enhanced to room temperature from the initial ~110 K. Such a behavior suggests an unconventional CDW feature, which is very rare, especially in such a three-dimensional bulk material. To understand this behavior, the powder x-ray diffraction and single crystal x-ray diffraction were carried out under pressure. The single crystal losses its intensity and the diffraction spots show obvious distortion above 5 GPa, suggesting strong lattice strain and a first-order phase transition. Clear lattice expansion along *a* axis was observed during the CDW transition in both powder and single crystal diffraction. At current stage, we did not observe diffraction from CDW superstructure beyond the abnormal lattice expansion, leading to the failure of resolving the exact crystal structure in the CDW phase. Hence, there are several open questions to be answered in near future: how does the CDW form in this new phase? Why the CDW temperature is enhanced? In this case, more delicate structure characterization is required, such as electron transmission spectroscopy or nano-beam synchrotron single crystal x-ray diffraction and even laser heating assisted x-ray diffraction; and systematic theoretical calculation based on the exact structure model is required to understand the unconventional CDW behavior. Anyhow, the Kagome metal BaFe$_2$O$_9$ provides a promising platform to study the unconventional CDW and its underlying physics. It is also promising to study the artificial BaFe$_2$O$_9$ thin film or chemical doped bulk crystal with tensile strain, which should help to suppress the CDW and could be beneficial for potential superconductivity.

## Acknowledgements

The authors acknowledge financial support from the National Key R&D Program of China (Grant No. 2021YFA1400300) and the National Natural Science Foundation of China (Grant Nos. 12374050). Part of the experimental work was carried out at the Synergetic Extreme Condition User Facility ((SECUF, https://cstr.cn/31123.02.SECUF).
## Acknowledgements

The authors acknowledge financial support from the National Key R&D Program of China (Grant No. 2021YFA1400300) and the National Natural Science Foundation of China (Grant Nos. 12374050). Part of the experimental work was carried out at the Synergetic Extreme Condition User Facility ((SECUF, https://cstr.cn/31123.02.SECUF).

## Author contributions

F. Hong conceived the project. M.W. Ma synthesized the single crystals. L.C. Chen and X.H. Yu did the high-pressure transport measurement and x-ray diffraction. F. Hong wrote the manuscript with comments from other authors.